\documentclass[11pt,a4paper]{article}
\usepackage[margin=27mm,top=25mm,bottom=27mm]{geometry}
\usepackage{amsmath,amsthm,booktabs,tikz}
\usepackage[font=small,labelfont=bf,labelsep=period]{caption}
\usepackage{hyperref}
\definecolor{teal}{HTML}{1F5863}
\definecolor{clay}{HTML}{963B2A}
\hypersetup{colorlinks=true,linkcolor=teal,citecolor=teal,urlcolor=teal,
pdftitle={Tight bounds on the sensitivity of LZ77 under k edits},pdfauthor={Shunsuke Inenaga}}
\newtheorem{theorem}{Theorem}
\newtheorem{lemma}[theorem]{Lemma}
\theoremstyle{definition}
\newtheorem{definition}[theorem]{Definition}
\newcommand{\ed}{\operatorname{ed}}
\newcommand{\word}[1]{\texttt{#1}}
\newcommand{\sep}{\mathbin{|}}
\tikzset{example alignment/.style={x=5.8mm,y=5.4mm,font=\small}}
\newcommand{\alignmentRows}{%
  \foreach \c in {7,9,10,12,23}
    \fill[clay!11] (\c-.46,-.05) rectangle (\c+.46,1.34);
  \node[anchor=east,font=\scriptsize] at (.37,2.85){Column};
  \foreach \c in {1,...,23} \node[font=\scriptsize] at (\c,2.85){\c};
  \node[anchor=east] at (.37,1.02){$T$};
  \node[anchor=east] at (.37,.25){$T'$};
  \foreach \c/\upper/\lower in
    {1/a/a,2/b/b,3/a/a,4/c/c,5/a/a,6/b/b,7/d/x,8/b/b,9/{-}/b,10/{-}/u,11/e/e,12/f/{-},13/b/b,14/b/b,15/g/g,16/d/d,17/b/b,18/e/e,19/f/f,20/b/b,21/!/!,22/Z/Z,23/{-}/w}{
    \node[font=\ttfamily] at (\c,1.02){\upper};
    \node[font=\ttfamily] at (\c,.25){\lower};}
  \path (-2.10,0) rectangle (24.70,0);
  \node[anchor=east,font=\scriptsize] at (.37,2.03){Factor};
  \foreach \lo/\hi/\i in {1/1/1,2/2/2,3/4/3,5/7/4,8/11/5,12/12/6,13/14/7,15/15/8,16/21/9,22/22/10}{
    \topGroup{\lo}{\hi}{$f_{\i}$}}
}
\newcommand{\topGroup}[3]{%
  \draw[teal] (#1-.43,1.43)--(#1-.43,1.61)--(#2+.43,1.61)--(#2+.43,1.43);
  \node[font=\footnotesize] at ({(#1+#2)/2},2.03){#3};}
\newcommand{\lowGroup}[3]{
  \draw (#1-.43,#3+.17)--(#1-.43,#3)--(#2+.43,#3)--(#2+.43,#3+.17);}
\newcommand{\highGroup}[3]{
  \draw (#1-.43,#3-.17)--(#1-.43,#3)--(#2+.43,#3)--(#2+.43,#3-.17);}

\title{Tight bounds on the sensitivity of LZ77 under $k$ edits}
\author{Shunsuke Inenaga\\ \normalsize Department of Informatics, Kyushu University}
\date{}

\begin{document}
\maketitle
\begin{abstract}
Let $z(T)$ be the number of factors in the self-referencing LZ77 factorization of a string $T$. We prove that for every nonempty $T$ and every integer $k\ge2$, any string $T'$ obtained from $T$ by at most $k$ single-character insertions, deletions, or substitutions satisfies $z(T')\le 3z(T)+3k-6$. For substitutions, this improves the upper bound $3z(T)+4k$ of Bathie, Huber, Lagarde, and Zemmari [arXiv:2602.19649v2]. For every $k\ge2$, we construct strings $T,T'$ over an alphabet of size $O(k)$ at both Hamming distance and edit distance exactly $k$ such that $z(T')=3z(T)+3k-6$.
\end{abstract}

\section{Introduction}
In the Lempel--Ziv 77 factorization~\cite{zl}, a factor can copy a substring that starts earlier in the text. Changing one letter in that earlier occurrence can therefore change factor boundaries far to its right. How large can the resulting factorization be after $k$ edits at arbitrary positions? We consider LZ77 with self-references, so a copied substring may overlap its source. Write $z=z(T)$ and $z'=z(T')$ for the numbers of factors before and after editing. Each edit inserts, deletes, or substitutes one character. The edit distance $\ed(T,T')$ is the minimum number of such edits needed to transform $T$ into $T'$. The Hamming distance between two strings of equal length is the number of positions at which they differ.

Akagi, Funakoshi, and Inenaga~\cite{afi} proved that a single edit to a nonempty string gives $z'\le2z$. Iterating this bound for $k$ substitutions gives $z'\le 2^k z$. Bathie, Huber, Lagarde, and Zemmari~\cite{bhlz} proved $z'\le3z+4k$ for at most $k$ substitutions. They also constructed instances with just two substitutions and arbitrarily large $z$ for which $z'/z$ approaches three. Hence any uniform bound of the form $z'\le Az+O(k)$ must have $A\ge3$. With the coefficient of $z$ fixed at three, what is the smallest additive term for each $k$, and does allowing insertions and deletions increase it? We answer both questions by proving the following theorem.

\begin{theorem}\label{th:main}
Let $T$ be nonempty and let $k\ge2$ be an integer. For every string $T'$ obtainable from $T$ by at most $k$ single-character edits,
\begin{equation}\label{eq:main}
 z(T')\le3z(T)+3k-6.
\end{equation}
For every $k\ge2$, there exist strings $T,T'$ at both Hamming distance and edit distance exactly $k$ satisfying $z(T')=3z(T)+3k-6$.
\end{theorem}

Thus the smallest additive term valid for all three edit operations is $3k-6$ for each $k\ge2$, and substitutions alone suffice for equality. When $k=2$, the bound becomes $z'\le3z$. In Section~\ref{sec:lower}, two substitutions turn a string with two factors into one with six, attaining exactly the ratio of three approached by the earlier instances. Table~\ref{tab:comparison} compares this result with the previous upper and lower bounds.

\begin{table}[ht]
\centering\small
\caption{Worst-case bounds on the number $z'=z(T')$ of LZ77 factors after editing a nonempty string $T$ with $z=z(T)$ factors. In the second row, $0<\delta<1$ is fixed. The instances attaining equality in the last row use an alphabet of size $O(k)$ and have Hamming distance and edit distance exactly $k$.}\label{tab:comparison}
\begin{tabular}{@{}lll@{}}\toprule
Edits & Upper bound on $z'$ & $z'$ for constructed pairs $(T,T')$\\\midrule
One edit~\cite{afi} & $z'\le 2z$ & $z'=2z$\\
At most $k$ substitutions~\cite{bhlz} & $z'\le3z+4k$ & $z'\ge(1-\delta)(3z+k)$ for infinitely many $k$\\
At most $k$ edits, $k\ge2$ (this work) & $z'\le3z+3k-6$ & $z'=3z+3k-6$ for every $k\ge2$\\\bottomrule
\end{tabular}
\end{table}

\paragraph{Proof ideas.}
By the optimality of greedy LZ77 parsing, it suffices to show that there exists a (possibly non-greedy) LZ factorization of $T'$ with at most $3z(T)+3k-6$ factors. We partition the image of each original factor $f_i$ under an edit alignment. Each resulting piece, with its last letter omitted, is empty or occurs in $T'$ starting before the piece, as required by Lemma~\ref{lem:greedy}. Breaks in $T$, placed initially at internal edit positions, mark where copies may have been damaged. Between consecutive breaks, the substring up to its penultimate letter is intact or occurs in the preceding factor images. Let $e_i$ be the number of edits belonging to $f_i$ and $t_i$ the number of breaks inside its source, with $t_i=0$ for a single letter. Cutting the image at the corresponding breaks and after its edit columns gives at most $1+t_i+e_i$ pieces by Lemma~\ref{lem:step}(i).

Figure~\ref{fig:step-block} illustrates why these same $t_i$ breaks can then be removed. Except where $f_i$ is edited, its image supplies a fresh copy of the source with its last letter omitted. Lemma~\ref{lem:step}(ii) places at most two endpoint breaks and one further break per edit in the image. Every break is removed at most once. There are at most $k$ initial breaks and at most $2z+k$ new ones, so $\sum_i t_i\le 2z+2k$. Including trailing insertions, the number of pieces is at most $z+k+\sum_i t_i$, giving $z'\le3z+3k$.

Now let $z\ge3$. The first factor is a single letter and places no breaks. The second can copy only from position one, so its source breaks can be counted directly against edits in the first two factors. The last factor needs no update after its image is cut. Only the middle factors $f_3,\ldots,f_{z-1}$ need the allowance of two endpoint breaks each, which saves six and gives $z'\le3z+3k-6$. The cases $z\le2$ follow directly for $k\ge2$.

The instances in Section~\ref{sec:lower} attain $z(T')=3z(T)+3k-6$ for every $k\ge2$. For $k=2$, we substitute two letters in a single run. For $k\ge3$, the original string consists of two identical copies followed by a fresh letter. The second copy and that letter form a single factor. All $k$ substitutions occur in the first copy. They split factors there and change which substrings can be copied when the parse reaches the untouched second copy. Choosing the run lengths appropriately gives $z(T)=k+1$ and $z(T')=6k-3$. Earlier lower-bound constructions for LZ77 sensitivity likewise edit one of two copies. A single substitution in the first copy can double the number of LZ77 factors~\cite{afi}, while replacing the entire second copy gives a lower bound on the increase in encoded length~\cite{bl}. Another construction changes two letters in a header that serves as a common source for later factors, making the ratio $z(T')/z(T)$ arbitrarily close to three~\cite{bhlz}.

\paragraph{Related work.}
Our upper-bound proof can be viewed as an extension of the single-edit argument of Akagi, Funakoshi, and Inenaga~\cite{afi} to $k>1$. As in their proof, we partition the edited string into pieces, each containing at most one factor start because its prefix up to the penultimate letter occurs earlier or is empty. For multiple substitutions, Bathie, Huber, Lagarde, and Zemmari~\cite{bhlz} bound the number of factor starts in each original interval by tracking the decrease in the depth of a jump function on the substitution positions. We bound the number of pieces in the aligned images. For each middle factor, the breaks inside its source specify cuts in its image and can then be removed. The alignment lets us use this argument for insertions and deletions as well.

For at most $k$ single-character edits at arbitrary positions, Gibney, Jin, Kociumaka, and Thankachan~\cite{gjkt} give an $O((z+k)\log N)$ bound on the edited factorization size, where $N=\max\{2,|T|,|T'|\}$ (Fact~2.3 in the full version). They use the pair-based parsing, in which a copied factor ends at the last letter of its match. By Kosolobov and Shur~\cite{ks}, its number of factors on a nonempty string $S$ lies between $z(S)$ and $2z(S)-1$, so the same asymptotic bound holds under our convention.

Fujie, Shibata, Nakashima, and Inenaga~\cite[Table~1]{fsni} obtained matching bounds of order $n^{2/3}$ for the largest increase $z'-z$ caused by a single edit to a string of length $n$. Blocki, Lee, and Yepes-Garcia~\cite{bly} studied the sensitivity of the encoded length of LZ77 under a single edit. Blocki and Lee~\cite{bl} considered edits confined to one contiguous interval.

\section{Factors and edit alignments}\label{sec:pre}
Let $\Sigma$ be a finite alphabet. For a string $S$ over $\Sigma$, write $|S|$ for its length and $\varepsilon$ for the empty string. Number positions from one. The substring $S[i..j]$ includes both endpoints, with $S[i..i-1]=\varepsilon$. Juxtaposition denotes concatenation, and $a^r$ denotes $r$ copies of the letter $a$ for an integer $r\ge0$.

A nonempty substring $S[p..q]$ is \emph{admissible} if $S[p..q-1]$ is empty or occurs in $S$ starting strictly before $p$. An \emph{LZ factorization} of $S$ is any factorization of $S$ into admissible substrings.

The \emph{LZ77 factorization} is the greedy LZ factorization. At each position $p$ where a factor begins, it takes the longest admissible prefix of $S[p..|S|]$. Equivalently, let $X$ be the longest nonempty prefix of $S[p..|S|]$ that occurs in $S$ starting before $p$, or set $X=\varepsilon$ if no such prefix exists. The factor consists of $X$ followed by the next character if there is one. The earlier occurrence of $X$ may overlap position $p$. This coincides with the convention used by Bathie, Huber, Lagarde, and Zemmari~\cite{bhlz}. The number of factors in the LZ77 factorization of a string $S$ is denoted by $z(S)$. Let $z(\varepsilon)=0$. We use a vertical bar to separate consecutive factors. For example, $\word{aaaaab}$ parses as $\word{a}\sep\word{aaaab}$. No endmarker is implicit.

The following lemma restates the optimality of the exhaustive history proved by Lempel and Ziv~\cite[Theorem~1]{lz76}. We include a short proof for completeness.

\begin{lemma}[Greedy parsing is optimal]\label{lem:greedy}
If $S=P_1\cdots P_r$ is an LZ factorization, then $z(S)\le r$.
\end{lemma}

\begin{proof}
We show that at most one factor starts within each $P_j=S[a..b]$. This is clear if $a=b$. Otherwise $S[a..b-1]$ occurs starting at some $a'<a$. If a factor starts at $p$ with $a\le p\le b$, then $S[p..b-1]$ occurs starting at $a'+p-a<p$. So the factor copies at least $b-p$ letters and then takes one more letter if any is left. It thus ends at $b$ or later, and no other factor starts in $P_j$.
\end{proof}

\paragraph{Edit alignments.}
Let $T=f_1\cdots f_z$ be the LZ77 factorization of a nonempty string $T$ of length $n$, and let $\ed(T,T')\le k$. Then we can fix an alignment of $T$ and $T'$ with at most $k$ edits, that is, a sequence of columns $(\sigma,\tau)\in(\Sigma\cup\{\varepsilon\})^2\setminus\{(\varepsilon,\varepsilon)\}$ whose upper entries spell $T$ and whose lower entries spell $T'$. A column is a \emph{match} if $\sigma=\tau$ and an \emph{edit} otherwise. In a match, $\tau$ is the \emph{copy} of the letter $\sigma$ of $T$. Copies occur in $T'$ in the order of their originals. A substring $T[x..y]$ \emph{survives} if $T[x],\dots,T[y]$ lie in consecutive match columns; their copies then spell $T[x..y]$. The \emph{position} of an edit is the number of letters of $T$ in that column and the columns before it. Thus substituting or deleting $T[q]$, and inserting a letter between $T[q]$ and $T[q+1]$, are edits at position $q$.
\begin{itemize}
\item[(A1)] If no edit has its position in $[x,y]$, then $T[x..y]$ survives.
\end{itemize}
The \emph{image} $\hat f_i$ of $f_i=T[L..R]$ is formed by the lower entries of the columns after that of $T[L-1]$, or from the first column if $L=1$, up to that of $T[R]$. The edits in these columns \emph{belong to} $f_i$, and $e_i$ denotes their number. The columns after that of $T[n]$ hold $e_{z+1}$ insertions, which form $\hat f_{z+1}$. Thus
\[
T'=\hat f_1\cdots\hat f_z\hat f_{z+1},\qquad e_1+\dots+e_{z+1}\le k.
\]
\begin{itemize}
\item[(A2)] The edits belonging to $f_i=T[L..R]$ have positions in $[L-1,R]$, and every edit at a position in $[L,R-1]$ belongs to $f_i$.
\end{itemize}
Cutting a part of $T'$ \emph{after} a column splits it between the lower entries up to that column and those after it.

\paragraph{Running example.}
Figures~\ref{fig:alignment} and~\ref{fig:example} use the same alignment. The original string has ten factors:
\[
 T=\word{a}\sep\word{b}\sep\word{ac}\sep\word{abd}\sep\word{be}\sep\word{f}
   \sep\word{bb}\sep\word{g}\sep\word{dbefb!}\sep\word{Z}.
\]
Substitute $\word{x}$ for $T[7]=\word{d}$, insert $\word{b}$ and $\word{u}$ between $T[8]$ and $T[9]$, delete $T[10]=\word{f}$, and append $\word{w}$. Thus $T'=\word{abacabxbbuebbgdbefb!Zw}$, and the five edits are at positions $7$, $8$, $8$, $10$, and $20$. Both insertions belong to $f_5=\word{be}$, whose image is $\hat f_5=\word{bbue}$. The image of the deleted letter $f_6=\word{f}$ is empty, and $\hat f_{11}=\word{w}$.

\begin{figure}[htbp]
\centering
\begin{tikzpicture}[example alignment]
\alignmentRows
\node[anchor=east,font=\scriptsize] at (.37,-.91){Image};
\node[anchor=east,font=\scriptsize] at (.37,-1.65){String};
\foreach \lo/\hi/\i/\out in
  {1/1/1/{\word{a}},2/2/2/{\word{b}},3/4/3/{\word{ac}},5/7/4/{\word{abx}},8/11/5/{\word{bbue}},12/12/6/{$\varepsilon$},13/14/7/{\word{bb}},15/15/8/{\word{g}},16/21/9/{\word{dbefb!}},22/22/10/{\word{Z}},23/23/11/{\word{w}}}{
  \draw[teal] (\lo-.43,-.24)--(\lo-.43,-.41)--(\hi+.43,-.41)--(\hi+.43,-.24);
  \node[font=\footnotesize] at ({(\lo+\hi)/2},-.91){$\hat f_{\i}$};
  \node[font=\footnotesize] at ({(\lo+\hi)/2},-1.65){\out};}
\node[anchor=east,font=\scriptsize] at (.37,-2.47){Position};
\foreach \j/\p/\edit in {1/1/0,2/2/0,3/3/0,4/4/0,5/5/0,6/6/0,7/7/1,8/8/0,9/8/1,10/8/1,11/9/0,12/10/1,13/11/0,14/12/0,15/13/0,16/14/0,17/15/0,18/16/0,19/17/0,20/18/0,21/19/0,22/20/0,23/20/1}{
  \ifnum\edit=1 \node[font=\footnotesize\bfseries,text=clay] at (\j,-2.47){\p};
  \else \node[font=\footnotesize] at (\j,-2.47){\p};\fi}
\end{tikzpicture}
\caption{The running alignment. Shading identifies the five edits, and \word{-} denotes $\varepsilon$. The lower brackets give the images $\hat f_i$ and the strings they spell. The bottom row gives, for each column, the number of letters of $T$ in it and the columns before it. Bold entries are the positions of the edits.}
\label{fig:alignment}
\end{figure}

\section{The upper bound}\label{sec:upper}

\paragraph{Sources and breaks.}
If $f_i=T[L..R]$ has $L<R$, then $T[L..R-1]$ occurs in $T$ starting at $L-\delta_i$ for some shift $\delta_i\ge1$. Thus $T[q]=T[q+\delta_i]$ for $L-\delta_i\le q<R-\delta_i$. We call the interval $[L-\delta_i,R-\delta_i]$ the \emph{source} of $f_i$. Edits can damage the source in $T'$, and then $\hat f_i$ may need several pieces even if no edit belongs to $f_i$. In the running example, $f_9=\word{dbefb!}$ copies $\word{dbefb}$ from $T[7..11]$, which four edits break up in $T'$. No edit belongs to $f_9$, yet four factors of $T'$ start in $\hat f_9$. A \emph{break} is a mark placed between $T[q]$ and $T[q+1]$ for some $1\le q<n$. At most one break is present at each position at any given time. Placing a break at a position from which a break has been removed creates a new break. We represent a set of breaks by the set of their positions. A break at position $q$ lies \emph{inside} the source of $f_i$ if $L-\delta_i\le q<R-\delta_i$. A set $B$ of breaks divides $T$ into \emph{blocks} $T[c..d]$, one for each pair $c-1<d$ of consecutive elements of $B\cup\{0,n\}$.

\begin{definition}\label{def:sound}
Let $1\le i\le z$. A set $B$ of breaks is \emph{$i$-sound} if, for each block $T[c..d]$ of $B$, the string $T[c..d-1]$ is empty, survives, or occurs in $\hat f_1\cdots\hat f_i$.
\end{definition}

Thus the text of each block, except possibly its last letter, is available in $T'$, either intact or within the images handled so far. Each substring of $T[c..d-1]$ has the same property, and an $i$-sound set is also $j$-sound for every $j>i$.

\begin{lemma}[One factor]\label{lem:step}
Let $2\le i\le z$ and $f_i=T[L..R]$, and let $B$ be $(i-1)$-sound. If $L<R$, write $\delta=\delta_i$ and let $t$ be the number of breaks of $B$ inside the source of $f_i$. If $L=R$, let $t=0$.
\begin{enumerate}
\item[(i)] The image $\hat f_i$ can be cut into at most $1+t+e_i$ admissible substrings of $T'$.
\item[(ii)] Suppose $L<R$. Define
\[
 K=\bigl(\{L-\delta-1,R-\delta\}
    \cup\{q-\delta \mid q \text{ is the position of an edit belonging to } f_i\}\bigr)
    \cap[1,n-1].
\]
Remove the $t$ breaks inside the source. Then place a new break at each position in $K$ at which no break remains. The resulting set $B'=(B\setminus[L-\delta,R-\delta-1])\cup K$ is $i$-sound with $|B'|\le|B|-t+2+e_i$.
\end{enumerate}
\end{lemma}

\begin{figure}[!htb]
\centering
\begin{tikzpicture}[x=9.4mm,y=8.3mm,font=\small,
  rowlabel/.style={anchor=east,font=\footnotesize},
  boxline/.style={draw=black!60,line width=.45pt},
  brk/.style={teal,line width=1.25pt},
  gone/.style={black!50,line width=.8pt},
  src/.style={fill=teal!13},
  kept/.style={fill=black!8},
  enddot/.style={fill=teal},
  editdot/.style={fill=clay}]
\node[rowlabel] at (-.05,3.02){$B$};
\fill[src] (1.1,2.73) rectangle (12.6,3.31);
\node[above,font=\footnotesize] at (6.85,3.31){source $[L-\delta,R-\delta]$};
\draw[black!60] (.05,3.02)--(13.55,3.02);
\foreach \a in {.52,13.20} \draw[brk] (\a,2.79)--(\a,3.25);
\foreach \a in {2.6,6.4,11.3}
  \draw[gone] (\a-.13,2.89)--(\a+.13,3.15) (\a-.13,3.15)--(\a+.13,2.89);
\node[rowlabel] at (.95,1.65){$f_i$};
\node[anchor=north west,font=\scriptsize,text=black!70] at (10.25,1.27){$f_i$ is drawn shifted by $-\delta$};
\fill[kept] (4.0,1.32) rectangle (8.72,1.98);
\draw[boxline] (1.1,1.32) rectangle (12.6,1.98);
\foreach \a in {4.0,8.72,10.04} \draw[boxline] (\a,1.32)--(\a,1.98);
\node at (6.36,1.65){$T[c+\delta..d-1+\delta]$};
\node at (9.38,1.65){$T[d+\delta]$};
\node at (2.55,1.65){$\cdots$}; \node at (11.32,1.65){$\cdots$};
\node[anchor=south east,font=\footnotesize,text=clay] at (4.03,2.00){edit at $q$};
\foreach \a in {4.0,10.04}{
 \fill[editdot] (\a,1.98) circle (1.5pt);
 \draw[densely dashed,clay,->] (\a,1.26)--(\a,.58);}
\node[rowlabel] at (-.05,.17){$B'$};
\fill[kept] (4,-.14) rectangle (8.72,.48);
\draw[boxline] (4,-.14) rectangle (10.04,.48);
\draw[boxline] (8.72,-.14)--(8.72,.48);
\node at (6.36,.17){$T[c..d-1]$};
\node at (9.38,.17){$T[d]$};
\draw[black!60] (.05,.17)--(4,.17) (10.04,.17)--(13.55,.17);
\foreach \a in {.52,13.20} \draw[brk] (\a,-.06)--(\a,.40);
\foreach \a/\s in {1.1/enddot,4.0/editdot,10.04/editdot,12.6/enddot}{
  \draw[brk] (\a,-.19)--(\a,.50); \fill[\s] (\a,.50) circle (1.7pt);}
\node[below,font=\footnotesize] at (1.1,-.19){$L-\delta-1$};
\node[below,font=\footnotesize] at (4,-.19){$q-\delta=c-1$};
\node[below,font=\footnotesize] at (10.04,-.19){$d$};
\node[below,font=\footnotesize] at (12.6,-.19){$R-\delta$};
\end{tikzpicture}
\caption{An interior block $T[c..d]$ of $B'$ in Lemma~\ref{lem:step}(ii). Bars mark breaks after the indicated positions of $T$, and the shading in row $B$ is the source of $f_i$. The crosses are the $t=3$ breaks of $B$ inside the source, which are removed. In row $B'$, dots mark new breaks: teal ones at the ends of the source and clay ones reached by the dashed arrows from edits of $f_i$. An edit in the gray part of $f_i$ would put a break inside the block. Thus the gray part survives in $\hat f_i$ and spells $T[c..d-1]$. If $L-\delta=1$, the left end of the source is the start of $T$ and needs no break.}
\label{fig:step-block}
\end{figure}
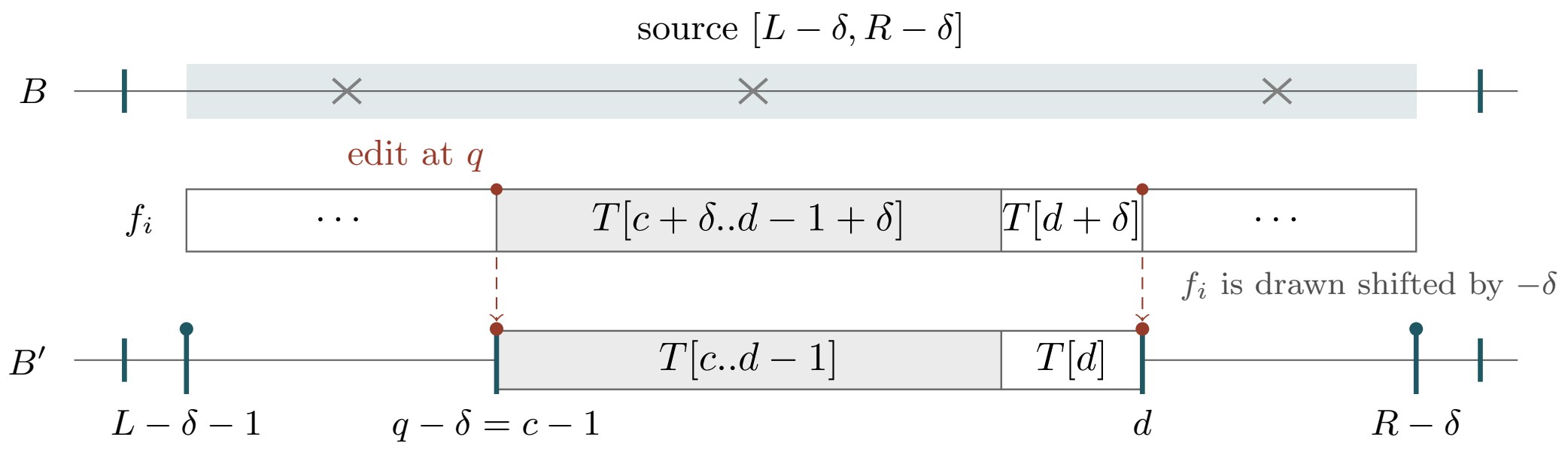

\begin{proof}
(i) Cut $\hat f_i$ after every edit and, if $L<R$, after the column of every $T[x]$ with $L\le x<R$ and $x-\delta\in B$. Discarding empty pieces leaves at most $1+t+e_i$ pieces. A piece from one column is a single letter. Let $P$ be a piece from two or more columns, as in Figure~\ref{fig:step-piece}. All of them but the last are matches, holding letters $T[x],\dots,T[x']$ with $L\le x\le x'<R$, so this case does not arise if $L=R$. Thus $T[x..x']$ survives, $P$ begins with the copy of $T[x]$, and $P$ without its last letter is a prefix of the copy of $T[x..x']$. None of $x-\delta,\dots,x'-\delta$ is in $B$, so $T[x-\delta..x'-\delta+1]$ lies within a block $T[c..d]$ of $B$, and $T[x..x']=T[x-\delta..x'-\delta]$ is a substring of $T[c..d-1]$. By $(i-1)$-soundness, $T[x-\delta..x'-\delta]$ either survives, and then its copy starts at the copy of $T[x-\delta]$, which precedes that of $T[x]$, or it occurs in $\hat f_1\cdots\hat f_{i-1}$. In both cases it occurs in $T'$ starting before $P$, so $P$ is admissible.

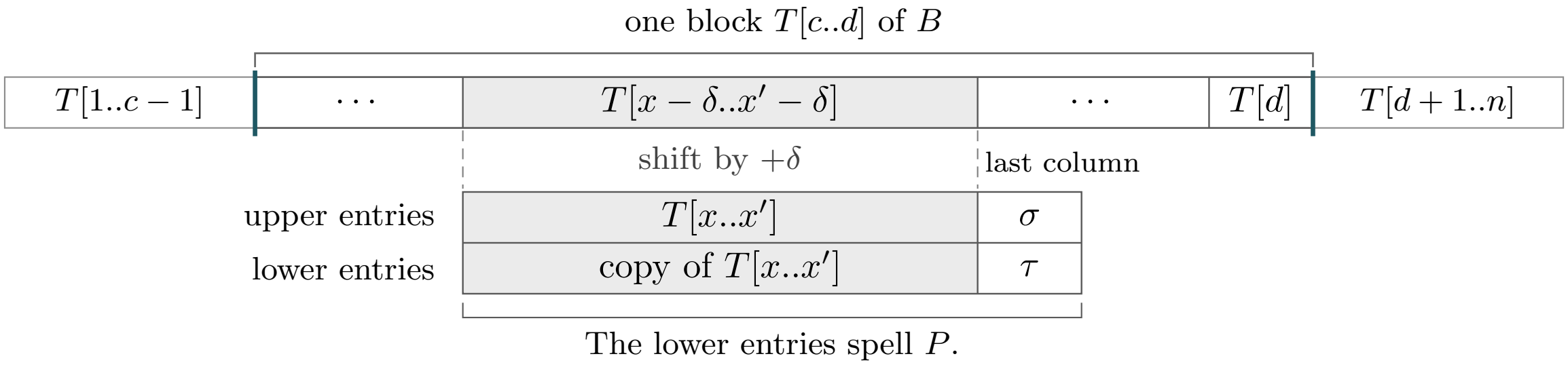
\begin{figure}[!htb]
\centering
\begin{tikzpicture}[x=9.2mm,y=8mm,font=\small,
  rowlabel/.style={anchor=east,font=\footnotesize},
  boxline/.style={draw=black!65,line width=.45pt},
  brk/.style={teal,line width=1.25pt},
  kept/.style={fill=black!8}]
\draw[black!60,line width=.5pt] (1.0,3.06)--(1.0,3.23)--(12.2,3.23)--(12.2,3.06);
\node[above,font=\footnotesize] at (6.6,3.23){one block $T[c..d]$ of $B$};
\fill[kept] (3.20,2.32) rectangle (8.65,2.94);
\draw[black!45,line width=.4pt] (-1.65,2.32) rectangle (14.85,2.94);
\draw[boxline] (1.0,2.32) rectangle (12.2,2.94);
\foreach \a in {3.20,8.65,11.1} \draw[boxline] (\a,2.32)--(\a,2.94);
\foreach \a in {1.0,12.2} \draw[brk] (\a,2.22)--(\a,3.02);
\node[font=\footnotesize] at (-.325,2.63){$T[1..c-1]$};
\node[font=\footnotesize] at (13.525,2.63){$T[d+1..n]$};
\node at (2.10,2.63){$\cdots$};
\node at (5.925,2.63){$T[x-\delta..x'-\delta]$};
\node at (9.875,2.63){$\cdots$};
\node at (11.65,2.63){$T[d]$};
\draw[densely dashed,black!50] (3.20,2.29)--(3.20,1.58);
\draw[densely dashed,black!50] (8.65,2.29)--(8.65,1.58);
\node[text=black!75,font=\footnotesize] at (5.925,1.92){shift by $+\delta$};
\node[anchor=west,font=\scriptsize,inner sep=0pt] at (8.72,1.91){last column};
\fill[kept] (3.20,.30) rectangle (8.65,1.54);
\draw[boxline] (3.20,.30) rectangle (9.75,1.54);
\draw[boxline] (3.20,.92)--(9.75,.92);
\draw[boxline] (8.65,.30)--(8.65,1.54);
\node[rowlabel] at (3.05,1.23){upper entries};
\node[rowlabel] at (3.05,.61){lower entries};
\node at (5.925,1.23){$T[x..x']$};
\node at (5.925,.61){copy of $T[x..x']$};
\node at (9.20,1.23){$\sigma$};
\node at (9.20,.61){$\tau$};
\draw[black!65] (3.20,.19)--(3.20,.01)--(9.75,.01)--(9.75,.19);
\node[below,font=\footnotesize] at (6.475,.01){The lower entries spell $P$.};
\end{tikzpicture}
\caption{Lemma~\ref{lem:step}(i). The two bars mark the boundaries of the block $T[c..d]$. The prefix and suffix on either side may be empty. Cuts in $\hat f_i$ at the shifted source breaks and after the edit columns give at most $1+t+e_i$ pieces. For the piece $P$ shown, the gray match columns spell a substring of $T[c..d-1]$. By $(i-1)$-soundness it has an occurrence in $T'$ starting before $P$. Each entry of the last column $(\sigma,\tau)$ is a letter or $\varepsilon$. Whether or not $\tau$ is empty, $P$ without its last letter is a prefix of the gray string.}
\label{fig:step-piece}
\end{figure}

(ii) Let $I=[L-\delta-1,R-\delta]$. By (A2), $K\subseteq I$, and $R-\delta<n$. Hence $B'\setminus I=B\setminus I$ and $|B'|\le|B|-t+|K|\le|B|-t+2+e_i$. Since $L-\delta-1$ and $R-\delta$ lie in $B'\cup\{0\}$, every block of $B'$ lies within the source or is disjoint from it. Let $T[c..d]$ be a block of $B'$ with $c<d$, so that no break of $B'$ lies in $[c,d-1]$. If the block is disjoint from the source, then $[c,d-1]$ is disjoint from $I$. Since $B'\setminus I=B\setminus I$, no break of $B$ lies in $[c,d-1]$ either. The block then lies within a block $T[c'..d']$ of $B$, and $T[c..d-1]$ is a substring of $T[c'..d'-1]$. Now suppose the block lies within the source. Figure~\ref{fig:step-block} shows how its prefix $T[c..d-1]$ corresponds to $T[c+\delta..d-1+\delta]$ in $f_i$. Since $[c+\delta,d-1+\delta]\subseteq[L,R-1]$, every edit at a position in this interval belongs to $f_i$ by (A2). No edit belonging to $f_i$ has its position $q$ in this interval, because $q-\delta\in K$ would put a break in $[c,d-1]$. Hence $T[c+\delta..d-1+\delta]$ survives by (A1). Its copy lies in $\hat f_i$ and spells $T[c..d-1]$.
\end{proof}

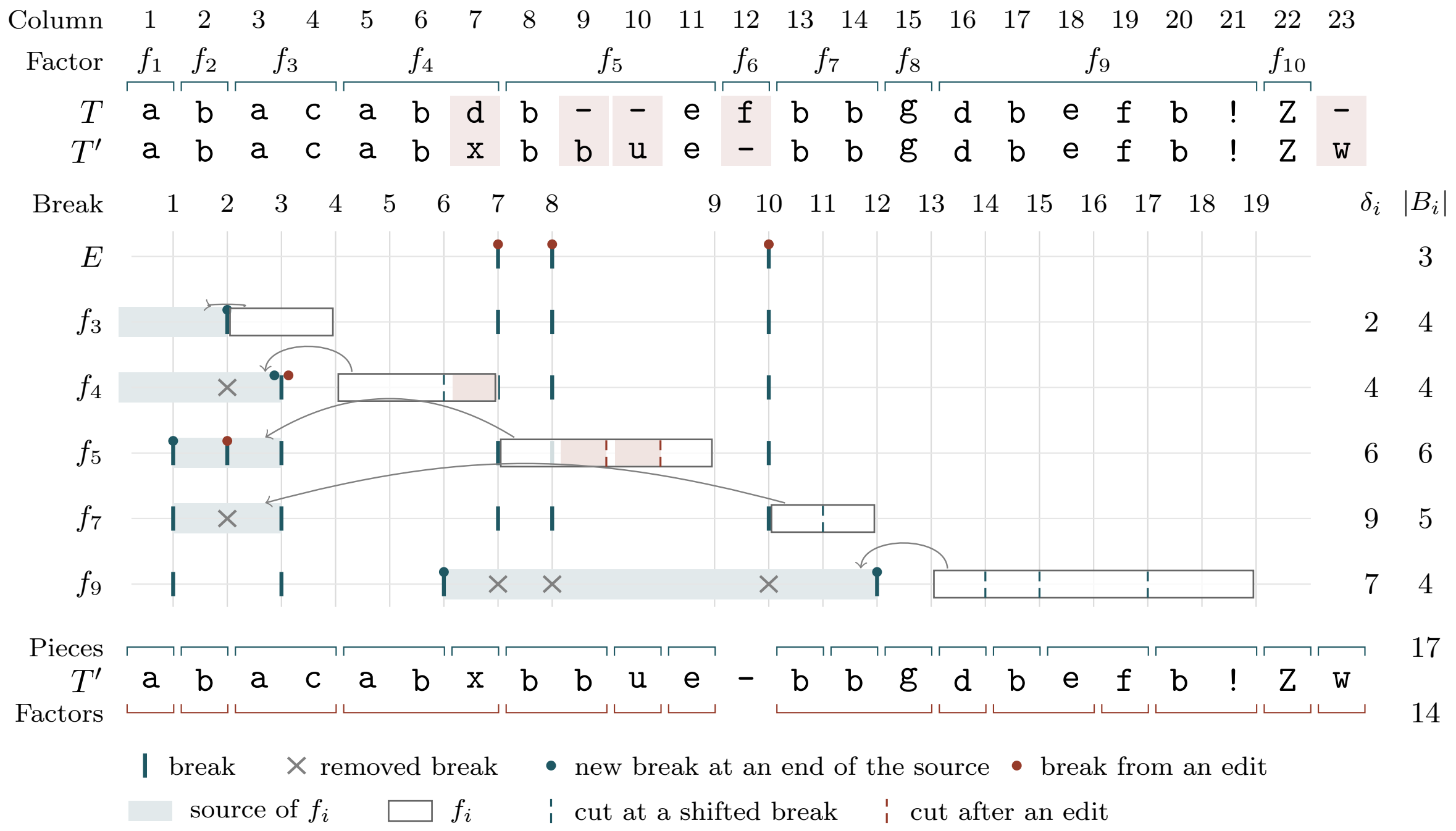
\begin{figure}[t]
\centering
\begin{tikzpicture}[example alignment,
  brk/.style={teal,line width=1.2pt},
  gone/.style={black!50,line width=.8pt},
  src/.style={fill=teal!13},
  fac/.style={draw=black!60,line width=.5pt},
  cut/.style={teal,densely dashed,line width=.6pt},
  ecut/.style={clay,densely dashed,line width=.6pt},
  enddot/.style={fill=teal},
  editdot/.style={fill=clay}]
\alignmentRows
\node[anchor=east,font=\scriptsize] at (.37,-.80){Break};
\foreach \p in {1,...,19}{\pgfmathtruncatemacro{\col}{\p+2*(\p>8)}
  \draw[black!14] (\col+.42,-1.35)--(\col+.42,-8.8);
  \node[font=\scriptsize] at (\col+.42,-.80){\p};}
\node[font=\scriptsize] at (23.55,-.80){$\delta_i$};
\node[font=\scriptsize] at (24.55,-.80){$|B_i|$};
\node[anchor=east,font=\footnotesize] at (.37,-1.85){$E$};
\draw[black!10] (.65,-1.85)--(22.43,-1.85);
\draw[brk] (7.42,-2.09)--(7.42,-1.61); \fill[editdot] (7.42,-1.61) circle[radius=1.5pt];
\draw[brk] (8.42,-2.09)--(8.42,-1.61); \fill[editdot] (8.42,-1.61) circle[radius=1.5pt];
\draw[brk] (12.42,-2.09)--(12.42,-1.61); \fill[editdot] (12.42,-1.61) circle[radius=1.5pt];
\node[font=\footnotesize] at (24.55,-1.85){3};
\node[anchor=east,font=\footnotesize] at (.37,-3.15){$f_{3}$};
\fill[src] (0.42,-3.45) rectangle (2.42,-2.85);
\draw[black!10] (.65,-3.15)--(22.43,-3.15);
\draw[brk] (7.42,-3.39)--(7.42,-2.91);
\draw[brk] (8.42,-3.39)--(8.42,-2.91);
\draw[brk] (12.42,-3.39)--(12.42,-2.91);
\draw[brk] (2.42,-3.39)--(2.42,-2.91);
\fill[enddot] (2.42,-2.91) circle[radius=1.5pt];
\fill[white,fill opacity=.8] (2.47,-3.42) rectangle (4.37,-2.88);
\draw[fac] (2.47,-3.42) rectangle (4.37,-2.88);
\draw[black!50,->,line width=.45pt] (2.72,-2.84) to[bend right=168] (2.12,-2.84);
\node[font=\footnotesize] at (23.55,-3.15){2};
\node[font=\footnotesize] at (24.55,-3.15){4};
\node[anchor=east,font=\footnotesize] at (.37,-4.45){$f_{4}$};
\fill[src] (0.42,-4.75) rectangle (3.42,-4.15);
\draw[black!10] (.65,-4.45)--(22.43,-4.45);
\draw[brk] (7.42,-4.69)--(7.42,-4.21);
\draw[brk] (8.42,-4.69)--(8.42,-4.21);
\draw[brk] (12.42,-4.69)--(12.42,-4.21);
\draw[gone] (2.26,-4.61)--(2.58,-4.29) (2.26,-4.29)--(2.58,-4.61);
\draw[brk] (3.42,-4.69)--(3.42,-4.21);
\fill[enddot] (3.29,-4.21) circle[radius=1.5pt]; \fill[editdot] (3.55,-4.21) circle[radius=1.5pt];
\fill[white,fill opacity=.8] (4.47,-4.72) rectangle (7.37,-4.18);
\fill[clay!14] (6.58,-4.7) rectangle (7.42,-4.2);
\draw[fac] (4.47,-4.72) rectangle (7.37,-4.18);
\draw[cut] (6.42,-4.72)--(6.42,-4.18);
\draw[black!50,->,line width=.45pt] (4.72,-4.14) to[bend right=87] (3.12,-4.14);
\node[font=\footnotesize] at (23.55,-4.45){4};
\node[font=\footnotesize] at (24.55,-4.45){4};
\node[anchor=east,font=\footnotesize] at (.37,-5.75){$f_{5}$};
\fill[src] (1.42,-6.05) rectangle (3.42,-5.45);
\draw[black!10] (.65,-5.75)--(22.43,-5.75);
\draw[brk] (3.42,-5.99)--(3.42,-5.51);
\draw[brk] (7.42,-5.99)--(7.42,-5.51);
\draw[brk] (8.42,-5.99)--(8.42,-5.51);
\draw[brk] (12.42,-5.99)--(12.42,-5.51);
\draw[brk] (1.42,-5.99)--(1.42,-5.51);
\fill[enddot] (1.42,-5.51) circle[radius=1.5pt];
\draw[brk] (2.42,-5.99)--(2.42,-5.51);
\fill[editdot] (2.42,-5.51) circle[radius=1.5pt];
\fill[white,fill opacity=.8] (7.47,-6.02) rectangle (11.37,-5.48);
\fill[clay!14] (8.58,-6) rectangle (9.42,-5.5);
\fill[clay!14] (9.58,-6) rectangle (10.42,-5.5);
\draw[fac] (7.47,-6.02) rectangle (11.37,-5.48);
\draw[ecut] (9.42,-6.02)--(9.42,-5.48);
\draw[ecut] (10.42,-6.02)--(10.42,-5.48);
\draw[black!50,->,line width=.45pt] (7.72,-5.44) to[bend right=32] (3.12,-5.44);
\node[font=\footnotesize] at (23.55,-5.75){6};
\node[font=\footnotesize] at (24.55,-5.75){6};
\node[anchor=east,font=\footnotesize] at (.37,-7.05){$f_{7}$};
\fill[src] (1.42,-7.35) rectangle (3.42,-6.75);
\draw[black!10] (.65,-7.05)--(22.43,-7.05);
\draw[brk] (1.42,-7.29)--(1.42,-6.81);
\draw[brk] (3.42,-7.29)--(3.42,-6.81);
\draw[brk] (7.42,-7.29)--(7.42,-6.81);
\draw[brk] (8.42,-7.29)--(8.42,-6.81);
\draw[brk] (12.42,-7.29)--(12.42,-6.81);
\draw[gone] (2.26,-7.21)--(2.58,-6.89) (2.26,-6.89)--(2.58,-7.21);
\fill[white,fill opacity=.8] (12.47,-7.32) rectangle (14.37,-6.78);
\draw[fac] (12.47,-7.32) rectangle (14.37,-6.78);
\draw[cut] (13.42,-7.32)--(13.42,-6.78);
\draw[black!50,->,line width=.45pt] (12.72,-6.74) to[bend right=15] (3.12,-6.74);
\node[font=\footnotesize] at (23.55,-7.05){9};
\node[font=\footnotesize] at (24.55,-7.05){5};
\node[anchor=east,font=\footnotesize] at (.37,-8.35){$f_{9}$};
\fill[src] (6.42,-8.65) rectangle (14.42,-8.05);
\draw[black!10] (.65,-8.35)--(22.43,-8.35);
\draw[brk] (1.42,-8.59)--(1.42,-8.11);
\draw[brk] (3.42,-8.59)--(3.42,-8.11);
\draw[gone] (7.26,-8.51)--(7.58,-8.19) (7.26,-8.19)--(7.58,-8.51);
\draw[gone] (8.26,-8.51)--(8.58,-8.19) (8.26,-8.19)--(8.58,-8.51);
\draw[gone] (12.26,-8.51)--(12.58,-8.19) (12.26,-8.19)--(12.58,-8.51);
\draw[brk] (6.42,-8.59)--(6.42,-8.11);
\fill[enddot] (6.42,-8.11) circle[radius=1.5pt];
\draw[brk] (14.42,-8.59)--(14.42,-8.11);
\fill[enddot] (14.42,-8.11) circle[radius=1.5pt];
\fill[white,fill opacity=.8] (15.47,-8.62) rectangle (21.37,-8.08);
\draw[fac] (15.47,-8.62) rectangle (21.37,-8.08);
\draw[cut] (16.42,-8.62)--(16.42,-8.08);
\draw[cut] (17.42,-8.62)--(17.42,-8.08);
\draw[cut] (19.42,-8.62)--(19.42,-8.08);
\draw[black!50,->,line width=.45pt] (15.72,-8.04) to[bend right=87] (14.12,-8.04);
\node[font=\footnotesize] at (23.55,-8.35){7};
\node[font=\footnotesize] at (24.55,-8.35){4};
\node[anchor=east,font=\scriptsize] at (.37,-9.6){Pieces};
\node[anchor=east] at (.37,-10.25){$T'$};
\node[anchor=east,font=\scriptsize] at (.37,-10.9){Factors};
\foreach \c/\l in {1/a,2/b,3/a,4/c,5/a,6/b,7/x,8/b,9/b,10/u,11/e,12/{-},13/b,14/b,15/g,16/d,17/b,18/e,19/f,20/b,21/!,22/Z,23/w}
  \node[font=\ttfamily] at (\c,-10.25){\l};
\begin{scope}[teal]
\foreach \lo/\hi in {1/1,2/2,3/4,5/6,7/7,8/9,10/10,11/11,13/13,14/14,15/15,16/16,17/17,18/19,20/21,22/22,23/23}
  {\highGroup{\lo}{\hi}{-9.6}}
\end{scope}
\begin{scope}[clay]
\foreach \lo/\hi in {1/1,2/2,3/4,5/7,8/9,10/10,11/11,13/15,16/16,17/18,19/19,20/21,22/22,23/23}
  {\lowGroup{\lo}{\hi}{-10.9}}
\end{scope}
\node[font=\footnotesize] at (24.55,-9.6){17};
\node[font=\footnotesize] at (24.55,-10.9){14};
\begin{scope}[yshift=-11.95*5.4mm,font=\scriptsize]
\draw[brk] (.9,-.24)--(.9,.24); \node[anchor=west] at (1.1,0){break};
\draw[gone] (3.54,-.16)--(3.86,.16) (3.54,.16)--(3.86,-.16); \node[anchor=west] at (3.9,0){removed break};
\fill[enddot] (8.4,0) circle[radius=1.5pt]; \node[anchor=west] at (8.6,0){new break at an end of the source};
\fill[editdot] (17.0,0) circle[radius=1.5pt]; \node[anchor=west] at (17.2,0){break from an edit};
\fill[src] (.6,-1.1) rectangle (1.4,-.7); \node[anchor=west] at (1.5,-.9){source of $f_i$};
\draw[fac] (5.4,-1.1) rectangle (6.2,-.7); \node[anchor=west] at (6.3,-.9){$f_i$};
\draw[cut] (8.4,-1.14)--(8.4,-.66); \node[anchor=west] at (8.6,-.9){cut at a shifted break};
\draw[ecut] (14.6,-1.14)--(14.6,-.66); \node[anchor=west] at (14.8,-.9){cut after an edit};
\end{scope}
\pgfresetboundingbox
\path (-1.7,-13.25) rectangle (25.0,3.15);
\end{tikzpicture}
\caption{The proof of Theorem~\ref{th:main} on the running alignment, choosing the leftmost occurrence as the source for each factor. The numbers below the alignment are positions in $T$, and a bar under $q$ is a break after $T[q]$. Row $E$ shows $B_2=E$, and the row of $f_i$ shows $B_i$. The rows for the single-letter factors $f_6$ and $f_8$ are omitted since $B_6=B_5$ and $B_8=B_7$. In each row, the arrow points from $f_i$ to its source, crosses mark the breaks removed inside the source, and dots mark new breaks. The dashed lines in a box are the cuts of $\hat f_i$, and each teal one lies at a removed break shifted by $\delta_i$. Breaks of $B_i$ inside a box are drawn faintly, since they do not cut $\hat f_i$. The bottom rows show the $17$ pieces of $T'$ produced by the proof and the $14$ LZ77 factors of $T'$. Each factor starts in a different piece, as in the proof of Lemma~\ref{lem:greedy}.}
\label{fig:example}
\end{figure}

\begin{proof}[Proof of the upper bound in Theorem~\ref{th:main}]
Let $E$ be the set of edit positions in $[1,n-1]$, so $|E|\le k$. For every block $T[c..d]$ of $E$, $T[c..d-1]$ is empty or survives by (A1). Thus $E$ is $i$-sound for every $1\le i\le z$.

First we bound the number of pieces for any choice of sound break sets. Put $t_1=0$. The first factor is a single letter, so $\hat f_1$ can be cut into at most $1+e_1$ letters. For $2\le i\le z$, use an $(i-1)$-sound set $B_{i-1}$ and let $t_i$ be the number of its breaks inside the source of $f_i$, or zero if $|f_i|=1$. By Lemma~\ref{lem:step}(i), $\hat f_i$ gives at most $1+t_i+e_i$ admissible pieces. The $e_{z+1}$ trailing insertions give one letter each. Since $\sum_{i=1}^{z+1}e_i\le k$, Lemma~\ref{lem:greedy} yields
\begin{equation}\label{eq:count}
 z(T')\le z+k+\sum_{i=1}^z t_i.
\end{equation}

\begin{samepage}
\emph{The bound $3z+3k$.}
Start with $B_1=E$. For $2\le i\le z$, form $B_i$ from $B_{i-1}$ by Lemma~\ref{lem:step}(ii) when $|f_i|\ge2$, and keep $B_i=B_{i-1}$ otherwise. In either case $B_i$ is $i$-sound. The $t_i$ breaks counted in~\eqref{eq:count} are exactly the breaks removed at $f_i$ (Figure~\ref{fig:step-block}), and at most $2+e_i$ new breaks are placed. Since every break is removed at most once,
\[
 \sum_{i=1}^z t_i\le |E|+\sum_{i=2}^z(2+e_i)\le k+(2z+k).
\]
Substituting in~\eqref{eq:count} gives $z(T')\le3z+3k$.\par
\end{samepage}

\emph{The refinement to $3z+3k-6$.}
For $z\ge3$, take $B_1=B_2=E$ and use the preceding update only for the middle factors $f_i$ with $3\le i<z$. The set $E$ is already $2$-sound, and the last image can be cut using $B_{z-1}$ without forming $B_z$. Inequality~\eqref{eq:count} holds with the $t_i$ for these choices. If $f_2=T[2..R]$ has $R>2$, its source starts at position one, so $t_2$ counts the positions of $E$ in $[1,R-2]$. Every edit at such a position belongs to $f_1$ or $f_2$. Hence $t_2\le e_1+e_2$, also when $R=2$.

The middle factors remove $\sum_{i=3}^{z-1}t_i$ breaks in total. Since $t_z\le |B_{z-1}|$, the breaks counted at the last factor are also covered by the initial breaks and those placed at middle factors. Thus
\[
 \sum_{i=3}^z t_i\le |E|+\sum_{i=3}^{z-1}(2+e_i).
\]
Adding the bound for $t_2$ gives $\sum_{i=1}^z t_i\le |E|+2(z-3)+\sum_{i=1}^{z-1}e_i\le 2(z-3)+2k$. By~\eqref{eq:count}, $z(T')\le3z+3k-6$.

If $z=2$, take $B_1=E$. The same bound $t_2\le e_1+e_2\le k$ gives $z(T')\le2+2k\le3k$. If $z=1$, then $t_1=0$ gives $z(T')\le1+k\le3k-3$. Both final inequalities use $k\ge2$.
\end{proof}

\paragraph{The running example.}
Figure~\ref{fig:example} uses the break sets for the sharper bound, choosing the leftmost occurrence for each source. Here $E=\{7,8,10\}$, since the trailing insertion is at position $n=20$. The factor $f_3=\word{ac}$ copies $\word{a}$ from $T[1]$, so its source is $[1,2]$, and $B_3=E\cup\{2\}$. The break $2$ lies inside the source $[1,3]$ of $f_4=\word{abd}$, so $\hat f_4=\word{abx}$ is cut into $\word{ab}$ and $\word{x}$. The substitution in $f_4$, at position $7$, gives the break $7-4=3$, which is also the right end of the source. The two insertions in $f_5$ both give position $8-6=2$, so they place a single new break there. The earlier break at this position was removed for $f_4$. For $f_9=\word{dbefb!}$, the source is $[7,12]$, and the breaks $7$, $8$, and $10$ inside it cut $\hat f_9$ into $\word{d}$, $\word{b}$, $\word{ef}$, and $\word{b!}$. They are then replaced by $6$ and $12$. Altogether $T'$ is cut into $17$ pieces, and each of its $14$ LZ77 factors starts in a different piece.

\section{Matching lower bound}\label{sec:lower}
For every $k\ge2$, substitutions alone suffice to attain the upper bound of Section~\ref{sec:upper}. We construct strings $T,T'$ at Hamming and edit distance $k$ with $z(T')=3z(T)+3k-6$.

\paragraph{Idea.}
A \emph{run} is a maximal substring $a^r$ with $r\ge1$. We call a run a \emph{record} if it is longer than every earlier run of $a$. The substitutions create records that start at factor boundaries. If the longest earlier run has length $M$ (zero for the first run), the first factor copies $M$ letters and takes one more $a$. The lengths and following letters in our construction satisfy Lemma~\ref{lem:runs}(b), so the remaining $a$'s and the following letter form a second factor. The edits also destroy the earlier occurrence of a long block, forcing new factor boundaries in its unedited copy.

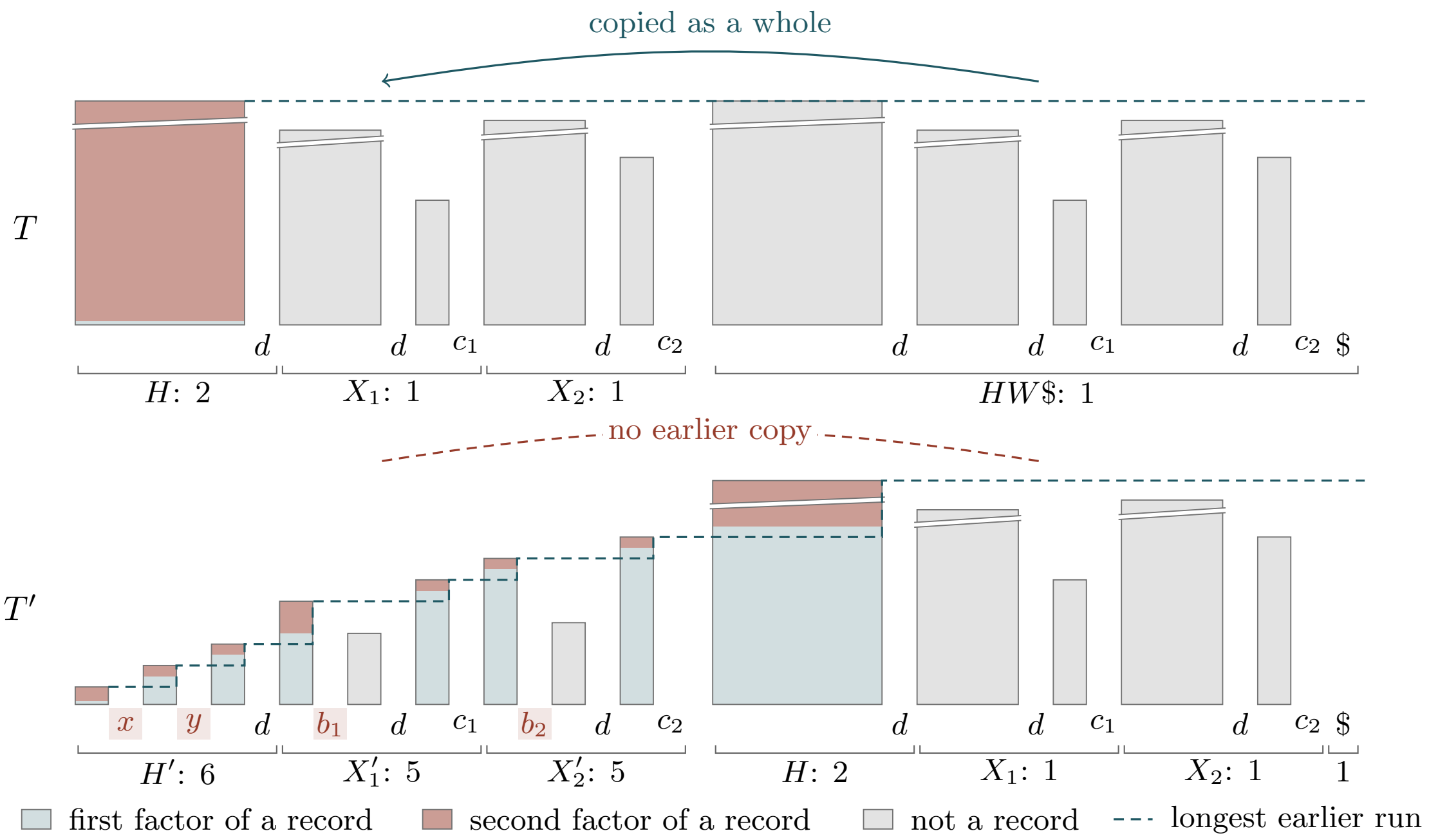
\begin{figure}[t]
\centering
\begin{tikzpicture}[x=1cm,y=1cm,font=\footnotesize,
  copied/.style={fill=teal!20},
  extra/.style={fill=clay!50},
  plain/.style={fill=black!11},
  edge/.style={draw=black!55,line width=.35pt},
  maxline/.style={teal,densely dashed,line width=.6pt},
  edit/.style={text=clay,font=\footnotesize\bfseries}]
\newcommand{\rec}[4]{\fill[copied] (#1,0) rectangle (#2,#3);
  \fill[extra] (#1,#3) rectangle (#2,#4); \draw[edge] (#1,0) rectangle (#2,#4);}
\newcommand{\run}[3]{\fill[plain] (#1,0) rectangle (#2,#3); \draw[edge] (#1,0) rectangle (#2,#3);}
\newcommand{\brk}[3]{\fill[white] (#1-.03,#3-.04) -- (#2+.03,#3+.03) -- (#2+.03,#3+.08) -- (#1-.03,#3+.01) -- cycle;
  \draw[black!55,line width=.35pt] (#1-.03,#3-.04) -- (#2+.03,#3+.03) (#1-.03,#3+.01) -- (#2+.03,#3+.08);}
\newcommand{\lett}[2]{\node at (#1,-.2){$#2$};}
\newcommand{\edt}[2]{\fill[clay!12] (#1-.17,-.36) rectangle (#1+.17,-.04); \node[edit] at (#1,-.2){$#2$};}
\newcommand{\reg}[3]{\draw[black!60] (#1+.03,-.42)--(#1+.03,-.50)--(#2-.03,-.50)--(#2-.03,-.42);
  \node at ({(#1+#2)/2},-.69){#3};}
\newcommand{\secondblocks}{%
  \run{8.65}{9.69}{2.00} \brk{8.65}{9.69}{1.86} \lett{9.87}{d} \run{10.05}{10.39}{1.28} \lett{10.57}{c_1}
  \run{10.75}{11.79}{2.10} \brk{10.75}{11.79}{1.94} \lett{11.97}{d} \run{12.15}{12.49}{1.72} \lett{12.67}{c_2}
  \lett{13.03}{\$}}

\begin{scope}[yshift=3.9cm]
\node[anchor=east,font=\small] at (-.25,1.0){$T$};
\rec{0}{1.74}{.04}{2.30} \brk{0}{1.74}{2.05} \lett{1.92}{d}
\run{2.10}{3.14}{2.00} \brk{2.10}{3.14}{1.86} \lett{3.32}{d} \run{3.50}{3.84}{1.28} \lett{4.02}{c_1}
\run{4.20}{5.24}{2.10} \brk{4.20}{5.24}{1.94} \lett{5.42}{d} \run{5.60}{5.94}{1.72} \lett{6.12}{c_2}
\run{6.55}{8.29}{2.30} \brk{6.55}{8.29}{2.05} \lett{8.47}{d}
\secondblocks
\draw[maxline] (1.74,2.30)--(13.25,2.30);
\reg{0}{2.1}{$H$: 2} \reg{2.1}{4.2}{$X_1$: 1} \reg{4.2}{6.3}{$X_2$: 1} \reg{6.55}{13.21}{$HW\$$: 1}
\draw[teal,->,line width=.6pt] (9.9,2.5) to[bend right=9] node[above,text=teal]{copied as a whole} (3.15,2.5);
\end{scope}

\node[anchor=east,font=\small] at (-.25,1.0){$T'$};
\rec{0}{.34}{.04}{.18} \edt{.52}{x}
\rec{.70}{1.04}{.29}{.40} \edt{1.22}{y}
\rec{1.40}{1.74}{.51}{.62} \lett{1.92}{d}
\rec{2.10}{2.44}{.73}{1.06} \edt{2.62}{b_1}
\run{2.80}{3.14}{.73} \lett{3.32}{d}
\rec{3.50}{3.84}{1.17}{1.28} \lett{4.02}{c_1}
\rec{4.20}{4.54}{1.39}{1.50} \edt{4.72}{b_2}
\run{4.90}{5.24}{.84} \lett{5.42}{d}
\rec{5.60}{5.94}{1.61}{1.72} \lett{6.12}{c_2}
\rec{6.55}{8.29}{1.83}{2.30} \brk{6.55}{8.29}{2.05} \lett{8.47}{d}
\secondblocks
\draw[maxline] (.34,.18)--(1.04,.18)--(1.04,.40)--(1.74,.40)--(1.74,.62)--(2.44,.62)--(2.44,1.06)
  --(3.84,1.06)--(3.84,1.28)--(4.54,1.28)--(4.54,1.50)--(5.94,1.50)--(5.94,1.72)--(8.29,1.72)--(8.29,2.30)--(13.25,2.30);
\reg{0}{2.1}{$H'$: 6} \reg{2.1}{4.2}{$X_1'$: 5} \reg{4.2}{6.3}{$X_2'$: 5}
\reg{6.55}{8.65}{$H$: 2} \reg{8.65}{10.75}{$X_1$: 1} \reg{10.75}{12.85}{$X_2$: 1} \reg{12.85}{13.21}{1}
\draw[clay,densely dashed,line width=.6pt] (9.9,2.5) to[bend right=9] node[pos=.5,fill=white,inner sep=1.5pt,text=clay]{no earlier copy} (3.15,2.5);

\begin{scope}[xshift=-.55cm,yshift=-1.2cm]
\fill[copied] (0,-.08) rectangle (.3,.12); \draw[edge] (0,-.08) rectangle (.3,.12);
\node[anchor=west] at (.35,.02){first factor of a record};
\fill[extra] (4.12,-.08) rectangle (4.42,.12); \draw[edge] (4.12,-.08) rectangle (4.42,.12);
\node[anchor=west] at (4.47,.02){second factor of a record};
\fill[plain] (8.65,-.08) rectangle (8.95,.12); \draw[edge] (8.65,-.08) rectangle (8.95,.12);
\node[anchor=west] at (9.00,.02){not a record};
\draw[maxline] (11.22,.02)--(11.62,.02); \node[anchor=west] at (11.67,.02){longest earlier run};
\end{scope}
\end{tikzpicture}
\caption{The construction for $k=4$. Bars are the runs of $a$ in order, and taller bars are longer runs. The heights are schematic, and a slanted cut marks a bar much taller than drawn. The letters between the runs appear below, with the edits shaded, and the dashed line is the length of the longest earlier run. For each record shown, the two colors mark the letters of the run in the first and second factors. The second factor also includes the letter following the run. The numbers count the factors in each part, $5$ in $T$ and $21=3\cdot5+3\cdot4-6$ in $T'$.}\label{fig:idea}
\end{figure}

For $k\ge3$, the original string $T=(HW)(HW)\$$ consists of two copies of a string $HW$, made of a header $H=a^Ud$ and $m=k-2$ blocks, followed by a fresh letter $\$$ (Figure~\ref{fig:idea}). Each block forms one factor with a source around the $d$ of $H$. The second $HW$ and the final $\$$ form one factor. In $T'=(H'W')(HW)\$$, two substitutions cut the run of the header into three, and one substitution in each block cuts its first run into two. Most runs of the first copy become records. In the unedited suffix $HW\$$, the header now takes two factors, each block takes one, and $\$$ forms a final one-letter factor. The gain $z(T')-z(T)=5m+6$ equals $3k+2m$, exactly the three factors per edit and two per middle factor allowed by the proof in Section~\ref{sec:upper}.

\begin{lemma}[Runs]\label{lem:runs}
Let a factor of $S$ start at the first letter of a run $a^r$ followed by a letter $c$, and let $M$ be the maximum length of an earlier run, or $M=0$ if there is none. Call $a^sc$ \emph{new} if it does not occur in $S$ starting before the run.
\begin{enumerate}
\item[(a)] If $r\le M$ and $a^rc$ is new, then $a^rc$ is a factor.
\item[(b)] If $r\ge M+2$ and $a^{r-M-1}c$ is new, then $a^{M+1}$ and $a^{r-M-1}c$ are consecutive factors.
\end{enumerate}
\end{lemma}

\begin{proof}
Let the run start at $p$. If $p>1$, then $S[p-1]\ne a$, so an occurrence of $a^j$ that starts before $p$ also ends before $p$. It lies within an earlier run, so $j\le M$. If $M>0$, then $a^{\min(r,M)}$ occurs within an earlier run of length $M$. If $M=0$, there is no earlier occurrence of $a$. (a) Thus $a^r$ occurs starting before $p$, while $a^rc$ does not, so the factor is $a^rc$. (b) Here the longest previous match at $p$ is $a^M$, and the factor is $a^{M+1}$. The next factor starts at $p+M+1$ and copies the remaining $r-M-1$ letters $a$ from one position to the left. No occurrence of $a^{r-M-1}c$ starts within the run before it, since $c$ would then lie inside the run. As $a^{r-M-1}c$ is new, the factor is $a^{r-M-1}c$.
\end{proof}

The string $a^sc$ is new if $c$ is \emph{fresh}, that is, absent before the run, or if every earlier occurrence of $c$ is preceded by fewer than $s$ copies of $a$.

\begin{proof}[Proof of equality in Theorem~\ref{th:main}]
For $k=2$, let $T=a^{14}d$ and $T'=a^2x\,a^4y\,a^6d$ with letters $x,y$ absent from $T$, so $\ed(T,T')=2$. Lemma~\ref{lem:runs}(b) gives $T=a\sep a^{13}d$ and $T'=a\sep ax\sep a^3\sep ay\sep a^5\sep ad$, so $z(T')=6=3z(T)+3k-6$.

For $k\ge3$, let $m=k-2$ and use distinct letters $a,d,x,y,\$$ and $b_i,c_i$ for $1\le i\le m$. Set $\ell=6k$, $U=3\ell-4$, $\alpha_i=\ell+4i$, $\beta_i=\ell+i$, $A_i=\alpha_i+\beta_i+1$,
\[
 H=a^Ud,\quad H'=a^{\ell-4}x\,a^{\ell-2}y\,a^\ell d,\quad
 X_i=a^{A_i}d\,a^{\alpha_i+2}c_i,\quad
 X_i'=a^{\alpha_i}b_ia^{\beta_i}d\,a^{\alpha_i+2}c_i,
\]
$W=X_1\cdots X_m$, and $W'=X_1'\cdots X_m'$. Thus $H'$ and $X_i'$ arise from $H$ and $X_i$ by substituting $x,y$ and $b_i$ for copies of $a$. These $m+2$ letters are absent from $T$, so the Hamming and edit distances are both $m+2=k$. With $M_1=\ell$ and $M_i=\alpha_{i-1}+2$ for $i\ge2$, one checks directly that
\begin{equation}\label{eq:ineq}
\begin{gathered}
 \ell-4\ge2,\qquad \alpha_i\ge M_i+2,\qquad \beta_i<\alpha_i,\qquad \ell<\beta_1<\cdots<\beta_m,\\
 A_m<U,\qquad \alpha_m+2<A_1,\qquad \beta_m<U-\alpha_m-3.
\end{gathered}
\end{equation}

In $T$, Lemma~\ref{lem:runs}(b) with $M=0$ splits $H$ into $a\sep a^{U-1}d$. The $d$ of $H$ has $U$ copies of $a$ on its left and $A_1$ on its right. By~\eqref{eq:ineq}, every $X_i$ without its fresh last letter $c_i$ occurs with its $d$ aligned to this $d$, so $X_i$ is a factor. The second $HW$ copies the first, and $\$$ ends this factor. Hence $z(T)=m+3$.

In $T'$, apply Lemma~\ref{lem:runs} to the runs in the edited first copy $H'W'$ and in the following header $H$, in order. The first run starts at the beginning of $T'$. In each application the last factor ends with the letter following the run, so the next run starts at a factor boundary. This gives
\[
 H'=a\sep a^{\ell-5}x\sep a^{\ell-3}\sep ay\sep a^{\ell-1}\sep ad,\qquad
 X_i'=a^{M_i+1}\sep a^{\alpha_i-M_i-1}b_i\sep a^{\beta_i}d\sep a^{\alpha_i+1}\sep ac_i,
\]
and $a^{\alpha_m+3}\sep a^{U-\alpha_m-3}d$ for the second header. For the runs $a^{\beta_i}$ we have $M=\alpha_i>\beta_i$. Every other run in $H'W'H$ is a record of length at least $M+2$ by~\eqref{eq:ineq}. The letters $x,y,b_i,c_i$ and the $d$ ending $H'$ are fresh. Before the run $a^{\beta_i}$, every $d$ is preceded by a run of length $\ell$ or $\beta_j$ with $j<i$, so $a^{\beta_i}d$ is new. Before the second header, the longest run preceding a $d$ has length $\beta_m<U-\alpha_m-3$, so $a^{U-\alpha_m-3}d$ is also new.

In the second copy, each $X_i$ is again a single factor. Its prefix before $c_i$ occurs around the $d$ of the second header as in $T$, and $X_i$ itself does not occur earlier, because the only earlier $c_i$ ends $X_i'$, which differs from $X_i$ and has the same length. Finally, $\$$ is fresh, and
\[
 z(T')=6+5m+(2+m+1)=6m+9=3z(T)+3k-6.\qedhere
\]
\end{proof}

\begin{samepage}
\section*{Use of AI}
The author made extensive use of \mbox{OpenAI}'s GPT-6.2 Astra and Anthropic's Claude Opus~5.5 in developing the results and preparing the manuscript. Through discussions with GPT, the upper bound $z(T')\leq 3z(T)+3k-6$ and the matching lower-bound instance were discovered. The author then worked with Claude to simplify the upper-bound and lower-bound proofs. GPT and Claude were also used to write the text and to prepare the figures. The author verified all results and takes full responsibility for the content of this paper.

\end{samepage}

\end{document}